\pdfoutput=1

\documentclass[]{ceurart}

\usepackage{times}
\usepackage{latexsym}
\usepackage{graphicx}
\usepackage{amsmath}
\usepackage{booktabs}
\usepackage{tikz}
\usepackage{caption}
\usepackage{subcaption}
\usepackage{multicol}
\usepackage{graphicx}
\usepackage{booktabs}

\usepackage[utf8]{inputenc}
\usepackage{pgfplots}
\DeclareUnicodeCharacter{2212}{−}
\usepgfplotslibrary{groupplots,dateplot}
\usetikzlibrary{patterns,shapes.arrows}
\pgfplotsset{compat=newest}

\usepackage{stfloats}
\usepackage[T1]{fontenc}

\usepackage[utf8]{inputenc}

\usepackage{microtype}

\usepackage{todonotes}

\title{Understanding the Relation of User and News Representations
in Content-Based Neural News Recommendation}

\author{Lucas Möller}
\author{Sebastian Padó}
\address{Institute for Natural Language Processing, University of Stuttgart, Germany \\ \{lucas.moeller, pado\}@ims.uni-stuttgart.de}

\begin{document}

\copyrightyear{2021}
\copyrightclause{Copyright for this paper by its authors.
  Use permitted under Creative Commons License Attribution 4.0
  International (CC BY 4.0).}

\conference{Proceedings of the 10th International Workshop on News Recommendation and Analytics at SIGIR`22}

\begin{abstract}
A number of models for neural content-based news recommendation
have been proposed. However, there is limited
understanding of the relative importances of the three main
components of such systems (news encoder, user encoder, and scoring function) and the trade-offs involved.
In this paper, we assess the hypothesis that the most widely used
means of matching user and candidate news representations is not
expressive enough. We allow our system to model more complex relations between the two by assessing more expressive scoring functions. Across a wide range of baseline and established systems this results in consistent improvements of around 6 points in AUC. Our results also indicate a trade-off between the complexity of news encoder and scoring function: A fairly simple baseline model scores well above 68\% AUC on the MIND dataset and comes within 2 points of the published state-of-the-art, while requiring a fraction of the computational costs.
\end{abstract}
\maketitle

\section{Introduction}

\begin{figure}
    \centering
    \begin{tikzpicture}
        \node at (0,0) {
            \includegraphics[width=.9\linewidth]{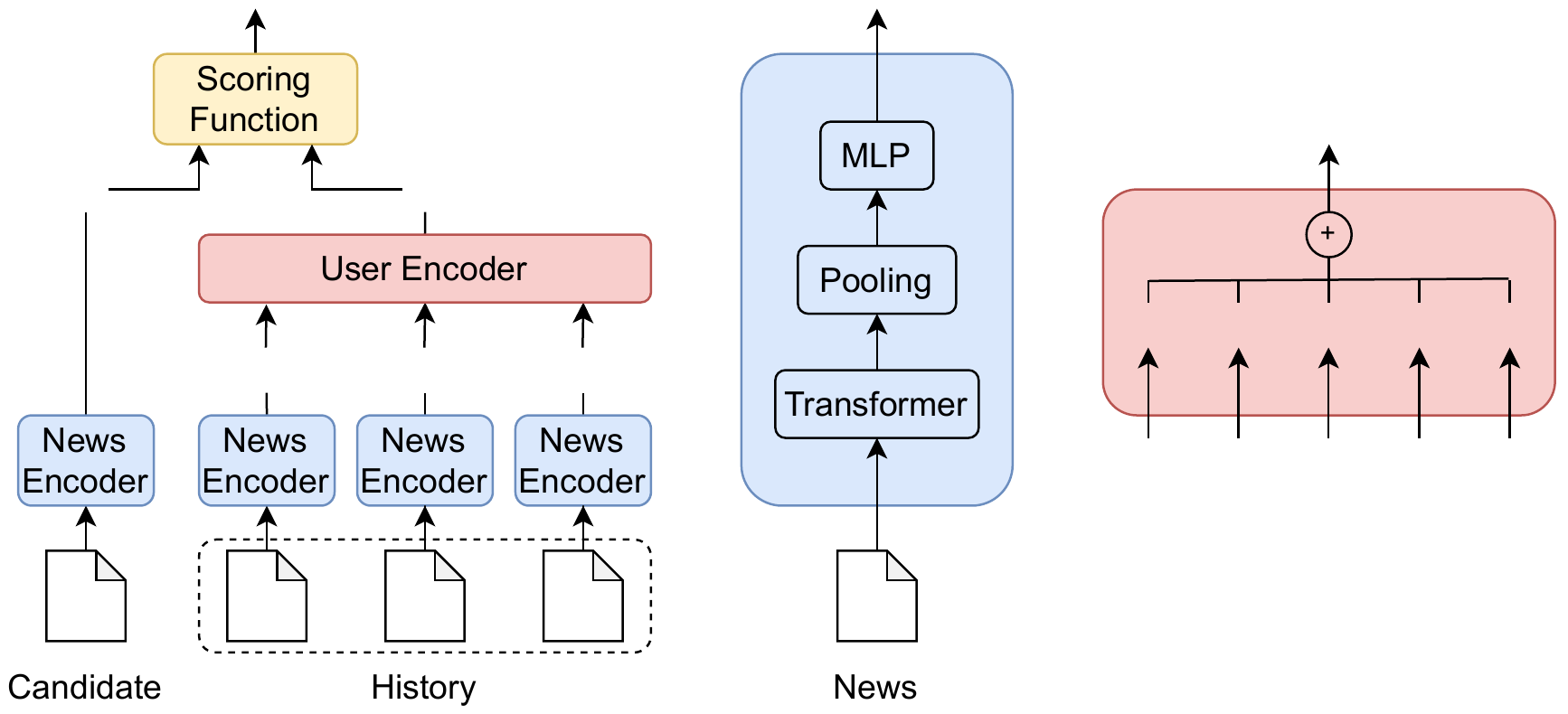}
        };
        \node at (-4, -3.5) {(a) Full Model};
        \node at (.8, -3.5) {(b) News Encoder};
        \node at (4.7, -3.5) {(c) User Encoder};
        \node at (-5.9, 1.45) {$\mathbf{c}$};
        \node at (-3.05, 1.45) {$\mathbf{u}$};
        \node at (-4.4, -.1) {$\mathbf{h}_1$};
        \node at (-3.05, -0.1) {$\mathbf{h}_2$};
        \node at (-1.7, -0.1) {$\mathbf{h}_3$};
        \node at (-4.45, 3.2) {$s$};
        \node at (0.8, 3.2) {$\mathbf{n}$};
        \node at (4.6, 2) {$\mathbf{u}$};
        \node at (3.1, 0.25) {$\lambda_1$};
        \node at (3.85, 0.25) {$\lambda_2$};
        \node at (4.6, 0.25) {$\lambda_3$};
        \node at (5.35, 0.25) {$\lambda_4$};
        \node at (6.1, 0.25) {$\lambda_5$};
        \node at (3.1, -1) {$\mathbf{h}_1$};
        \node at (3.85, -1) {$\mathbf{h}_2$};
        \node at (4.6, -1) {$\mathbf{h}_3$};
        \node at (5.35, -1) {$\mathbf{h}_4$};
        \node at (6.1, -1) {$\mathbf{h}_5$};
    \end{tikzpicture}
    \vspace{.5cm}
    \caption{Components of a typical neural NRS. The full model (a) consists of a news encoder (b), a user encoder (b) and a scoring function. Details are given in Section~\ref{sec:method}.}
    \label{fig:nrs}
\end{figure}
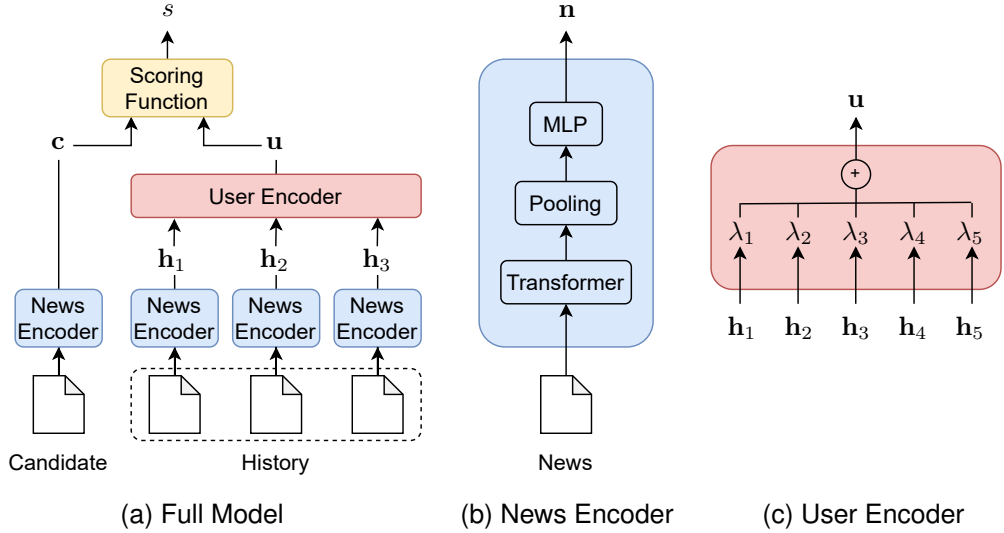

News recommender systems (NRS) guiding users to news items that are of interest to them are in widespread use \cite{yahoo, embedding_nrs, forbes, a_month_nrs, swissinfo}. Traditional approaches often relied on collaborative filtering and fought with a range of problems \cite{lian, content_cf, karimi, Lommatzsch, zgbek, RealtimeNRS}. In recent years \textit{neural content-based} approaches have successfully addressed many prior challenges \cite{deep_nrs, kumar, raza, survey_wu}.
Figure~\ref{fig:nrs} shows the architecture shared by many of these
systems. They typically consist of three components: (a), a \textit{news encoder} which maps individual news articles onto embeddings; (b), a \textit{user encoder} which produces user representations $\mathbf{u}$ as a function of their reading history $\mathbf{h}_t$; (c), a \textit{scoring function} that maps a pair of a candidate news
embedding $\mathbf{c}$ and a user representation $\mathbf{u}$ onto a
scalar score $s$.

In this paper,  we start from the observation that these three components
of the NRS architecture have received very different amounts of attention.
Regarding (a), the news encoder, there is a large amount of research, ranging from early applications of convolutional networks \cite{npa, wang} to the integration of additional features from topic models \cite{naml, topic_nrs,lstur}, or knowledge bases \cite{dkn, kred, knowldege_nrs} to modern pre-trained language models \cite{nrs_plm}. As for (b), the user encoder, some systems use recurrent models \cite{lstur, Lommatzsch, zhang} or graph representations \cite{hu_graph, user_as_graph, graph_enh}. However, state-of-the-art models obtain user representations from additive combinations of the user's reading history \cite{nrs_plm}. In a large majority of models, the scoring function (c) is then instantiated by an inner product \cite{npa, nrms, topic_nrs, naml, dkn, feedbacknet, feedrec, nrs_plm, hierec}.
We believe this is not expressive enough:
Given that, user representations are essentially averages of news embeddings, the use of a simple inner product entails that scores can only compare user and candidate news embeddings \textit{within} but not \textit{across} dimensions of the embedding space. 
This puts a large strain on the news encoder, since it needs to encode any such
relevant interactions explicitly in some dimension of the news
embedding.

Following up on this observation, we allow our system to model more complex relations between user and candidate news representations by systematically assessing more expressive scoring functions.
We evaluate these scoring functions in combination with a number of baseline and SOTA news recommender systems.\\
Our results show that a more complex, yet relatively simple, scoring function consistently results in a large improvement of the overall performance. Furthermore, it can obviate the need for a complex news encoder and still perform at the state of the art. In this
manner, we make a contribution to a better understanding of the roles and importances of the individual components in the general NRS architectures.

\section{Method} \label{sec:method} 

We implement a neural content-based NRS with the components shown in
Figure~\ref{fig:nrs}. Our news and user encoder closely follow previously published methods \cite{npa, nrs_plm}. The focus of our experiments is then on the scoring function, as it is this component that models the relation between candidate news and user representations.

\subsection{News Encoder}
\label{sec:news-encoder}

The content of a news article is typically represented by obtaining embeddings $e_i$ from a sequential model and subsequently pooling them into a fixed-length news vector
$\mathbf{n}$. We use a pre-trained transformer \cite{bert, attention} for embeddings and an additive attention mechanism from previous studies for pooling \citep{wang, dkn,naml, npa, nrms}:

\begin{equation} \label{eq:addatt}
\begin{split} 
    \mathbf{n} &= \sum_i \alpha_i \, \mathbf{e}_i \, ,\\
    \alpha_i &= \text{softmax} \left( \mathbf{q}^T \, \text{tanh} \left( W \, \mathbf{e}_j + \mathbf{b} \right) \right)_i
\end{split}
\end{equation}
    
\noindent The indexes $i$ and $j$ range over all token embeddings. $W$, $\mathbf{q}$ and $\mathbf{b}$ are parameters.
The pooled representation is further processed by two linear layers with ReLU activations. We initialize both the pooling mechanism and the linear layers randomly and train them together with the full model.

\subsection{User Encoder} \label{sec:user_encoder}

In line with previous studies \cite{dkn, npa, naml, kred}, we compute
user embeddings $\mathbf{u}$ from their reading histories. We combine the vector representation $\mathbf{h}_t$ of respective news by means of an additive attention mechanism analogous to the one used in the news encoder:
\begin{equation} \label{eq:user_rep}
    \mathbf{u} = \sum_t \lambda_t \, \mathbf{h}_t
\end{equation}

\noindent Here the index $t$ is over the last $T$ historic news a user has read, and $\lambda_t$ are computed analogous to $\alpha_i$ in Equation~\ref{eq:addatt}.

\subsection{Scoring Functions}
\label{sec:recommender-heads}
We now define a series of scoring functions to compute a score $s$ for
the match between a user embedding $\mathbf{u}$ and a candidate news
embedding $\mathbf{c}$. We focus on generalizations of the inner
product with increasing expressiveness.\\
As discussed above, a simple scoring function is an
inner product followed by a sigmoid transformation:
\begin{equation} \label{eq:dot_rec}
    s \left( \mathbf{u},\mathbf{c} \right) = \sigma \left( \mathbf{c}^T \mathbf{u} \right)
\end{equation}

\noindent However, its limitation becomes clear when we plug in Equation~\ref{eq:user_rep},

\begin{equation}
    \mathbf{c}^T \mathbf{u} = \sum_d c^d u^d = \sum_d \sum_t \lambda_t \, c^d h_t^d \, ,
\end{equation}
where $d$ indexes the dimension of the embedding space: Only dependencies within identical dimensions of history and candidate news are considered for the computation of $s$. The score cannot depend on dependencies across different dimensions of the feature space.

We can remove this limitation by defining a scoring function based on a bilinear form where off-diagonal entries in the matrix $A$ may capture interactions among different dimensions:
\begin{equation} \label{eq:bilin_rec}
    s = \sigma \left( \mathbf{c}^T A \, \mathbf{u} \right)
\end{equation}
When $A$ is treated as a parameter, the scoring function becomes a
learnable component.
If we add a bias $\mathbf{b}$ and an activation function $a$ we obtain a non-linear version:
\begin{equation} \label{eq:nonlin_rec}
    s = \sigma \left( \mathbf{c}^T a \left( A \, \mathbf{u} + \mathbf{b} \right) \right),
\end{equation}
Finally, we consider a two-layer MLP acting on the concatenation $\mathbf{u} || \mathbf{c}$ of the two representation vectors:
\begin{equation} \label{eq:mlp_rec}
    s = \sigma \left( W_2 \, a \left( W_1 \, [ {\mathbf{u} || \mathbf{c}} ] + \mathbf{b} \right) \right)
\end{equation}

\section{Experiments and Results}
\subsection{Experimental Setup}

\paragraph{Data.}
We carry out experiments on the widely used Microsoft News Dataset
(MIND, \citet{mind}) for news recommendation.
It consists of logs generated from one million
randomly sampled users over a period of six weeks and contains approximately 160k news items. Notably, this dataset also contains cold start sessions, i.e. sessions for which no user history exists.

\paragraph{Task and Model.}
We train a binary click-prediction classification task with a standard
cross-entropy objective. Each input is a news document presented to a user in a given session and the user's reading history up to this session. The output is whether the news document was clicked.\\
For a fair comparison among model architectures, we use a pre-trained roBERTa transformer \cite{roberta} to encode the news documents (cf.\ Section~\ref{sec:news-encoder}) throughout. We do not fine-tune the transformer, which permits us to train the full model on a single RTX 2070 GPU. 
\paragraph{Training.}
In each training iteration, we sample one clicked news and a number of
$K$ negatives from a given session. In accordance with previous studies, we
use $K\!=\!4$ and a batch size of $64$. We use the Adam optimizer with
a learning rate of $1e\!-\!4$. A maximum of the last $T\!=\!25$ news are
used from a user's reading history. The embedding dimensionality for
news and user vectors is set to $256$. All trainings run for five
epochs.
\paragraph{Evaluation.}
We evaluate our experiments with the standard ranking metrics
\textit{Area Under the Curve} (AUC), \textit{Mean Reciprocal Rank}
(MRR), and \textit{Normalized Discount Cumulative Gain} up to position
five (NDCG@5) and ten (NDCG@10). Cold start users receive random scores from a uniform distribution.\\
To test whether one model is significantly superior to another, we use del Barrio's test for stochastic dominance on the loss distributions of the respective models. This test is non-parametric and compares the percentile functions of two distributions \cite{delBarrio}. For a detailed explanation and the test's suitability for the evaluation of deep models we refer to the work by Dror et al. \cite{dror}. We choose a maximum violation level of $\epsilon\!=\!0.33$ and a significance level of $\alpha\!=\!0.01$.

\subsection{Experiment 1: Comparing Scoring Functions}

\begin{table}[tb]
	\centering
	\caption{Performances of different scoring functions and number of parameters for various models. Original model configurations are indicated by \textit{(orig.)}. Results for the best scoring function in each model and metric are underlined. Figure~\ref{fig:boxplots} visualizes these results. Refer to the text regarding significance of improvements.}
	\label{tab:results}
	\vspace{.3cm}
    \begin{tabular}{lccccr}
    \toprule
    \textbf{Model + Scoring Function} & \textbf{AUC} & \textbf{MRR} & \textbf{NDCG@5} & \textbf{NDCG@10} & \textbf{params}\\
    \cmidrule(r){1-1}
    \cmidrule(lr){2-5}
    \cmidrule(l){6-6}
    Base + inner & $62.59$ & $27.89$ & $29.69$ & $36.55$ & $526k$\\
    Base + bilinear & $67.50$ & $32.43$ & $35.58$ & $41.95$ & $591k$\\
    Base + nonlinear & \underline{$68.66$} & \underline{$32.66$} & \underline{$36.06$} & \underline{$42.45$} & $657k$\\
    Base + mlp & $67.99$ & $32.42$ & $35.71$ & $42.09$ & $592k$\\
    \midrule
    NPA + inner \textit{(orig.)} & $61.67$ & $27.62$ & $29.13$ & $36.16$ & $23.2M$\\
    NPA + bilinear & \underline{$68.23$} & $32.50$ & $35.78$ & $42.13$ & $23.2M$\\
    NPA + nonlinear & \underline{$68.23$} & \underline{$32.67$} & \underline{$36.02$} & \underline{$42.35$} & $23.3M$\\
    NPA + mlp & $68.08$ & $32.63$ & $35.92$ & $42.27$ & $23.2M$\\
    \midrule
    NAML + inner \textit{(orig.)}& $62.21$ & $26.67$ & $28.17$ & $35.35$ & $1.06M$\\
    NAML + bilinear & $67.89$ & $32.49$ & $35.81$ & $42.23$ & $1.13M$\\
    NAML + nonlinear & \underline{$67.90$} & \underline{$32.68$} & \underline{$35.99$} & \underline{$42.42$} & $1.20M$\\
    NAML + mlp & $67.89$ & $32.02$ & $35.29$ & $41.82$ & $1.13M$\\
    \midrule
    NRMS + inner \textit{(orig.)} & $68.57$ & \underline{$33.02$} & $36.20$ & \underline{$42.78$} & $3.15M$\\
    NRMS + bilinear & $68.40$ & $32.24$ & $35.58$ & $42.17$ & $3.22M$ \\
    NRMS + nonlinear & $68.74$ & $32.53$ & $35.96$ & $42.35$ & $3.28M$\\
    NRMS + mlp & \underline{$68.85$} & $32.85$ & \underline{$36.33$} & $42.75$ & $3.22M$\\
      \midrule
    NRMS ablation + inner & $63.82$ & $28.47$ & $30.57$ & $37.48$ & $2.89M$\\
    NRMS ablation + bilinear & \underline{$68.20$} & \underline{$32.28$} & \underline{$35.52$} & \underline{$42.15$} & $2.95M$\\
            \midrule
    Mean + inner & $58.89$ & $25.55$ & $27.13$ & $33.62$ & $263k$\\
    Mean + bilinear & $67.68$ & $32.51$ & $35.85$ & $42.15$ & $328k$\\
    Mean + nonlinear & \underline{$67.81$} & \underline{$32.52$} & \underline{$35.64$} & \underline{$42.02$} & $394k$\\
    Mean + mlp & $66.88$ & $32.08$ & $35.15$ & $41.40$ & $328k$\\
    \bottomrule
    \end{tabular}
\end{table}

In our first experiment we evaluate the performance of our Base model from Section~\ref{sec:method} in combination with all four scoring functions. The results are shown at the top of Table~\ref{tab:results} (first group of results).

We find a reasonable baseline performance of around 62.6\% AUC for the inner
product score. The bilinear scoring function clearly outperforms the
inner product by $5$ points in AUC (67.5\%). The non-linear scoring function
further improves the performance by one point to 68.7\%, which the MLP
cannot surpass (68.0\%). The improvement of the bilinear scoring function over the inner one and that of the nonlinear over the bilinear one are both
significant ($\epsilon\!=\!0$ and $\epsilon\!=\!0.29$).

Figure~\ref{fig:losses} (Base) shows the loss distributions of all four
models. Clearly, compared with the other models the
inner product has a distinctively lower peak at low values and a much
heavier tail towards higher values. It also has a sharp peak at
$log(0.5)\!\approx \! 0.7$, indicating the model is uncertain about a
substantial fraction of the data and places them right at the decision
boundary.

\subsection{Experiment 2: SOTA Models with Different Scoring Functions} \label{sec:exp2}

\begin{figure}
    \centering
    \includegraphics{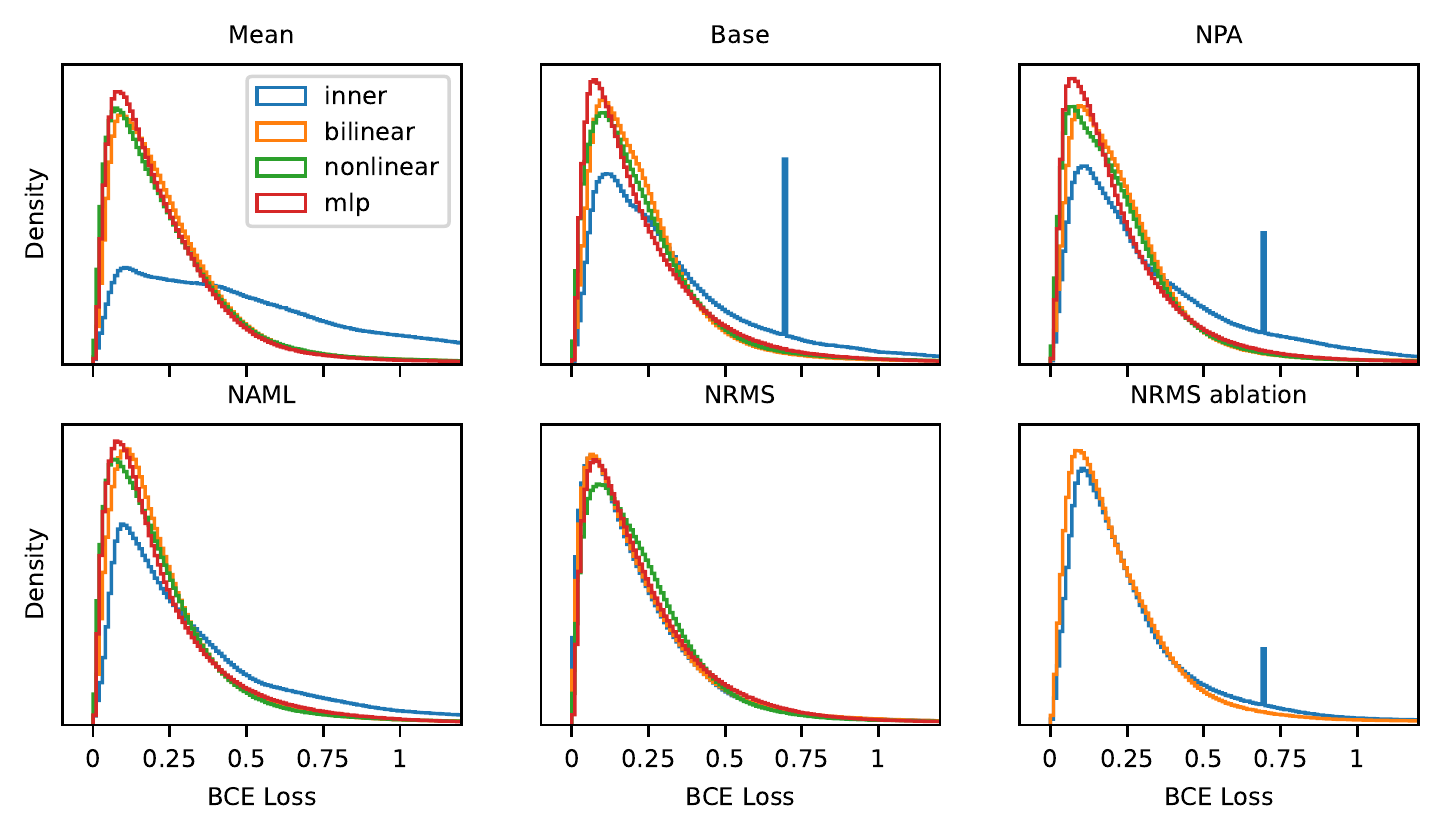}
    \vspace{-.15cm}
    \caption{Test loss distributions for all models and scoring functions.}
    \label{fig:losses}
    \vspace{-0.15cm}
\end{figure}

To investigate to what extent these patterns generalize beyond our
Base model, we now vary the scoring function in three state-of-the-art
NRS models: NPA personalizes the user encoder \cite{npa}, NAML
includes categorical and textual news features \cite{naml} and NRMS
applies multi-head attention in the user- and news-encoder
\cite{nrms}. All standardly use an inner product score. We complete
the set of models with a trivial Mean baseline which replaces the
attention mechanisms in the news and user encoders
(Equation~\ref{eq:addatt}, Equation~\ref{eq:user_rep}) with simple
averages.
The results are shown in the rest of
Table~\ref{tab:results}. Figure~\ref{fig:losses} visualizes the loss
distributions of all combinations.

Strikingly, for all models except NRMS the bilinear scoring function
largely outperforms the inner product. Moreover, the bilinear models
are strictly stochastically dominant over the inner product models
($\epsilon\!=\!0$), i.e. they outperform the latter at every
percentile of the loss distribution \cite{delBarrio, dror}.\\ Second,
our Base model from Experiment 1 performs within 0.2 percentage points
AUC of the best overall model (68.7\% vs. 68.9\%). Even the Mean model
in combination with the nonlinear scoring function comes to within 1
percentage point AUC of the best model (67.8\%). This is especially
interesting considering the poor performance of the Mean model in
combination with an inner product scoring function (58.9\%). By
changing only the scoring function, this trivial baseline can compete
with much more complex architectures.

In contrast, we cannot confirm a superiority of a nonlinear over a bilinear scoring function in this experiment. For the models tested here, the two show very similar results. Improvements, where present, are not significant.

The outlier model in this experiment is NRMS, for which the choice of
scoring function does not appear to matter much. We believe that this
is the case because NRMS is the only model that transforms the news
representations in a user's history before additively combining them
to a user embedding. To test whether this transformation plays an
important role, we remove it, obtaining the 'NRMS ablation' model. Indeed, this modification leads to a large drop in performance of almost 5 pp AUC to a level slightly above the NPA, NAML and Base model. When replacing the inner product with a bilinear score the performance recovers to
68.2\% AUC. This performance is not significantly worse than that of the original NRMS model ($\epsilon\!=\!0.45$).

\subsection{Meta Analysis} 
\begin{figure}
    \centering
    \includegraphics{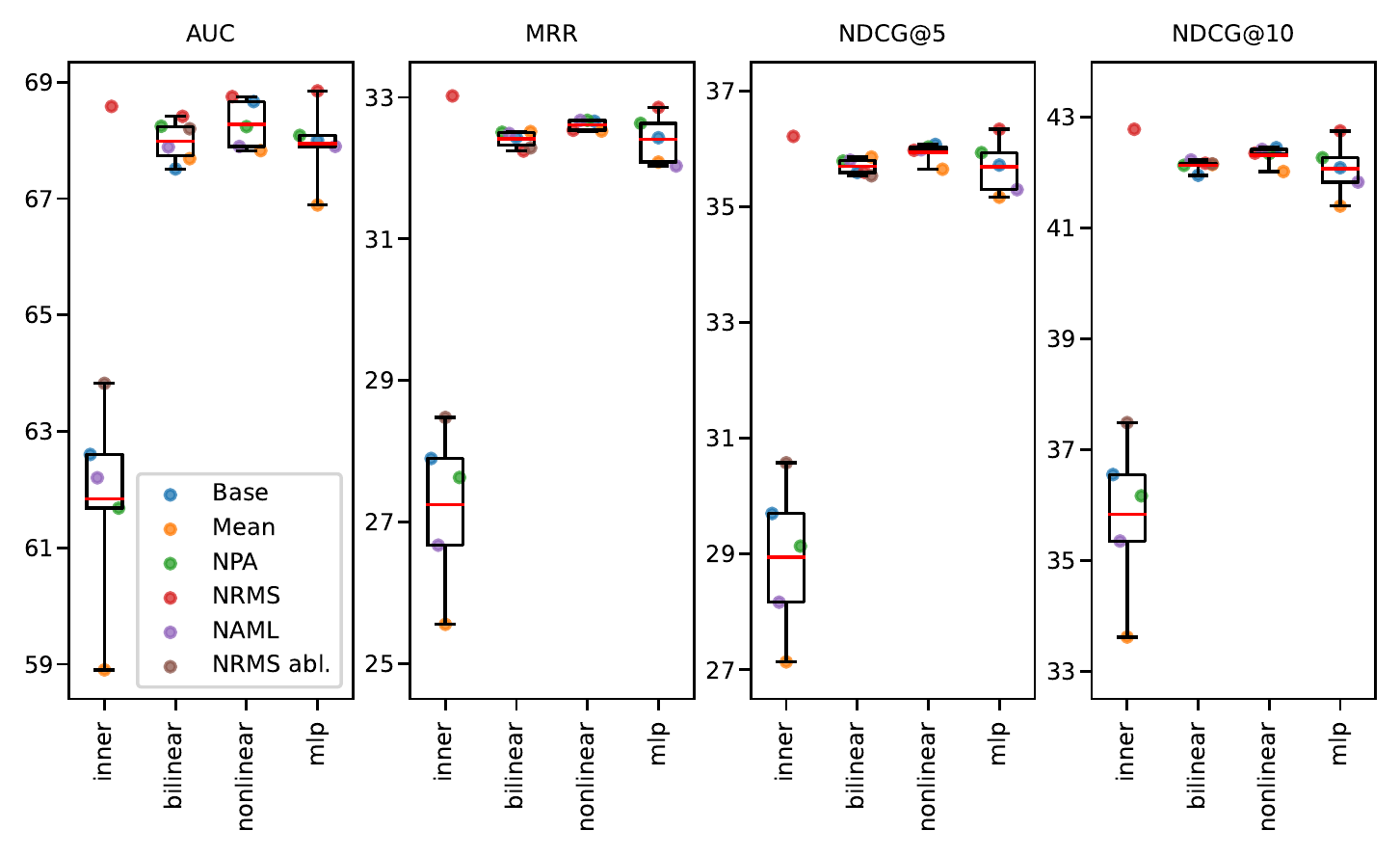}
    \caption{Boxplot visualizations of the four evaluation metrics for all combinations of models (legend) and scoring functions (x-axes). Red lines indicate means. Exact values can be found in Table~\ref{tab:results}}
    \label{fig:boxplots}
\end{figure}

We conclude by carrying out a meta analysis of the results across all combinations of scoring functions with the five implemented model architectures (Mean, Base, NPA, NAML, NRMS). Figure~\ref{fig:boxplots} visualizes the results from Tabel~\ref{tab:results} using boxplots.\\
The bilinear scoring function accounts for an average improvement of 6.1$\,\pm \,$1.5 points in AUC over a simple inner product\footnote{For
the NRMS model in combination with an inner product, we consider the
ablation described above.}. Very much in parallel, MRR increases by 5.2$\,\pm\,$1.1 pp, NDCG@5 by 6.7$\,\pm\,$1.3 and NDCG@10 by 6.3$\,\pm\,$1.3 points. On the contrary, there is hardly a difference between the bilinear, nonlinear or MLP scoring functions.

An interesting result is also that simple models (Mean and Base) in combination with more powerful scoring functions perform better than models with complex news encoders (NPA, NAML, NRMS ablation) combined with an inner product score. A more expressive scoring function appears to be able to compensate for complexity in other parts of the model.

Finally, Table~\ref{tab:results} also shows the number of parameters in
every model. NPA and NRMS are especially parameter hungry due to
their use of embedding and attention layers (20M and 3M, respectively). NAML has additional parameters for category embedding layers and a second news encoder for the abstract of the news. The Mean and Base models, on the other hand, only have some 100ks of parameters, meaning that they are cheaper and likely more robust to be learned.


\section{Conclusion}

In this paper, we have
dissected the relation of user and candidate news representation in content-based neural NRS, which is modeled by the scoring function.
On top of a range of baseline and SOTA models, we find a large
improvement of 6.2$\,\pm\,$1.4 points in AUC for moving from an inner
product to a bilinear form, but no further improvements for moving to
a nonlinear version or an MLP. These findings extend similar results on
collaborative approaches by \citet{rendle} to neural content-based
NRS.\\
By implementing a bilinear scoring function, a trivial baseline (Mean)
can almost reach a 1 pp AUC proximity of our best model, while having an order of magnitude less parameters. Our slightly
more complex Base model comes within a 2 pp AUC margin of the currently published state of the art \cite{nrs_plm}.\\
We achieve these results without fine-tuning the transformer backbone of the news encoder. Together with their small number of parameters, these models require relatively little computational costs. Thus, they can serve as conceptually simple and cheap, yet powerful baselines \cite{nrs_baselines}.

Overall, we conclude that representing users by means of an additive combination of historic news embeddings and subsequently using an inner
product to model the relation with candidate news is not sufficient ---
A more expressive relation between user and candidate news
representations can enhance the performance of NRS by a large margin
and can even compensate for complex news encoders.\\
We believe our study is a first step towards a systematic understanding of the importances of the individual components of NRS for their overall performance.

\bibliography{nrs}

\end{document}